\documentclass[11pt]{article}
\usepackage[utf8]{inputenc}
\usepackage{hanging}
\usepackage[margin=1in]{geometry}
\usepackage{indentfirst}
\usepackage{amsmath,amsfonts,amssymb,mathtools}
\usepackage{float}
\usepackage{soul}
\usepackage{comment}
\usepackage{graphicx}

\title{\bf Model Mixing Using Bayesian Additive \\Regression Trees}
\author{John C. Yannotty, Thomas J. Santner, Richard J. Furnstahl, and Matthew T. Pratola\hspace{.2cm}\\
    The Ohio State University}

\date{\today}

\newcommand{\iid}{\stackrel{iid}{\sim}}
\newcommand{\ind}{\stackrel{ind}{\sim}}

\newcommand{\ith}{\small{i\text{th}}}
\newcommand{\jth}{\small{j\text{th}}}
\newcommand{\pth}{\small{p\text{th}}}
\newcommand{\lth}{\small{l\text{th}}}
\newcommand{\kth}{\small{k\text{th}}}

\newcommand{\second}{\small{2\text{nd}}}
\newcommand{\fourth}{\small{4\text{th}}}
\newcommand{\sixth}{\small{6\text{th}}}

\newcommand{\M}{\mathcal{M}}

\newcommand{\D}{\mathcal{D}}

\newcommand{\R}{\mathbb{R}}

\newcommand{\betavec}{\boldsymbol\beta}
\newcommand{\muvec}{\boldsymbol\mu}
\newcommand{\wvec}{\boldsymbol w}
\newcommand{\fvec}{\boldsymbol f}
\newcommand{\fhatvec}{\boldsymbol{\hat{f}}}
\newcommand{\gvec}{\boldsymbol g}
\newcommand{\bvec}{\boldsymbol b}
\newcommand{\xvec}{\boldsymbol x}
\newcommand{\rvec}{\boldsymbol R}
\newcommand{\Fvec}{\boldsymbol F}
\newcommand{\Amat}{\boldsymbol A}
\newcommand{\onevec}{\boldsymbol 1}

\DeclareMathOperator*{\argmax}{argmax}

\usepackage{natbib}
\setcitestyle{authoryear,open={(}, close = {)}} 
\usepackage{wrapfig}
\usepackage[section]{placeins}

\usepackage{color}
\usepackage{soul} 

\begin{document}
\maketitle
\begin{abstract}
In modern computer experiment applications, one often encounters the situation where various models of a physical system are considered, each implemented as a simulator on a computer. An important question in such a setting is determining the best simulator, or the best combination of simulators, to use for prediction and inference. Bayesian model averaging (BMA) and stacking are two statistical approaches used to account for model uncertainty by aggregating a set of predictions through a simple linear combination or weighted average. Bayesian model mixing (BMM) extends these ideas to capture the localized behavior of each simulator by defining input-dependent weights. One possibility is to define the relationship between inputs and the weight functions using a flexible non-parametric model that learns the local strengths and weaknesses of each simulator. This paper proposes a BMM model based on Bayesian Additive Regression Trees (BART). The proposed methodology is applied to combine predictions from Effective Field Theories (EFTs) associated with a motivating nuclear physics application.
\end{abstract}

\noindent%


\section{Introduction}

In statistical learning problems, one often considers a set of plausible models, each designed to explain the system of interest. A common practice is to select a best performing model based on some pre-specified criteria. The ensuing inference for quantities of interest is then carried out using the selected model as if it were the true data generating mechanism. The resulting uncertainty quantification ignores any variability due to the underlying model structure \citep{draper1995assessment}. The misrepresentation of uncertainties associated with such quantities can ultimately lead to misguided interpretation or inappropriate decisions. Another shortcoming of the typical approach to modeling is that the resulting inference may strongly depend on the selection criteria. In other words, different sets of criteria could lead to noticeably different final models and inferential results. To account for such uncertainties, one may elect to combine information across the set of models in some manner.    

Any model set can be classified as $\M$-closed, $\M$-complete, or $\M$-open \citep{bernardo2009bayesian}. These three categories differ in their underlying assumptions regarding a true model, $\M_\dagger$, and its relation to the model set. 
The $\M$-closed setting assumes a mathematical representation of $\M_\dagger$ can be formulated and it is included in the model set. In this setting, model selection is appropriate because $\M_\dagger$ can be recovered from the set of models under consideration. The $\M$-complete setting also assumes it is possible to construct $\M_\dagger$, however it is not included in the model set. For example, an expression for $\M_\dagger$ may exist, however it may be computationally intensive or intractable compared to the models under consideration. The $\M$-open case assumes the true model may exist, however a lack of knowledge or resources prevents one from constructing its mathematical representation. Consequently, $\M_\dagger$ is excluded from the model set. 
This work is motivated by applications in nuclear physics which tend to fall within the $\M$-open class as the underlying truth regarding the physical system may not yet be understood. In such cases, one may desire to leverage the known information about the physical system which is contained in the model set along with experimental data to further understand the nuclear phenomena.  


Assume a set of $K$ models are considered when studying a particular system of interest. One approach to account for model uncertainty is to combine the information across these $K$ models. This may involve combining the individual point predictions or probability density functions from each model, usually in some additive manner. Traditional frequentist and Bayesian approaches utilize global weighting schemes, where each model is weighted by a value intended to reflect overall (global) model performance. For example, a classical global weighting scheme is Bayesian model averaging (BMA) \citep{bma_lr}, which combines the individual posterior densities from each model using a convex combination. The BMA weights are given by the individual posterior model probabilities, each which can be interpreted as the probability the  individual model is the true data generating one. Hence, BMA implicitly assumes the true model is contained within the model set, which renders this method inappropriate outside of the $\M$-closed setting \citep{bernardo2009bayesian}. More recent Bayesian global weighting schemes adopt a model stacking approach, where model weights are assigned to minimize a specified posterior expected loss. This decision theory viewpoint of global weighting can be used for combining point predictions \citep{le2017bayes} or probability densities \citep{cs}. Under some assumptions, stacking methods have been shown to be more appropriate for both the $\M$-open and $\M$-closed settings \citep{cs}.  

\begin{figure}[t]
    \centering
    \includegraphics[width = 0.9\textwidth, height = 0.37\textwidth]{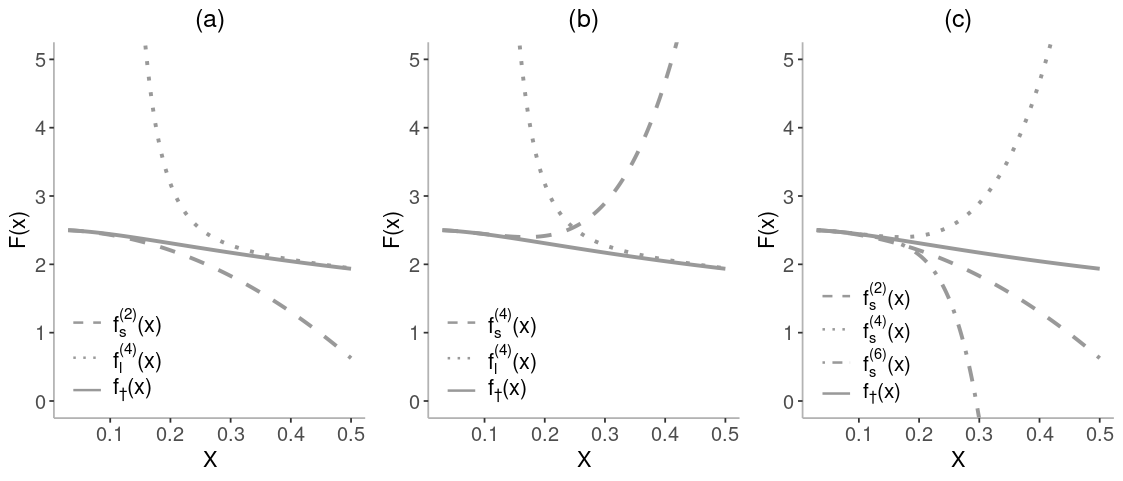}
    \caption{Three different EFT experimental settings.  Each panel displays the true physical system (solid) and the mean predictions from the EFTs under consideration (non-solid). }
    \label{fig:expansions}
\end{figure}

Though global weighting methods are effective, they still might lead to poor approximations of the true system when the individual model performance is localized. In such a case, one may wish to select a weighting scheme that reflects the localized characteristics of the models by constructing input-dependent weights. With input-dependent weights, one would expect an individual model to receive a higher weight in input regions where it exhibits strong predictive performance, while receiving a weight close to 0 in regions of poor performance. Localized weighting schemes are more appropriate for the $\M$-open or $\M$-complete settings where the true model may be better characterized as a localized mixture of the model set under consideration.

This work is motivated by problems in nuclear physics modeled using a technique known as Effective Field Theory (EFT)~\citep{burgess_2020,doi:10.1142/8619,Georgi:1994qn}. EFTs are designed to perform well in a particular subregion(s) of the input domain, yet diverge in the rest of the input domain. Prototypes of such models are the weak and strong coupling finite-order expansions for the partition function of the zero-dimensional $\phi^4$ theory presented by \cite{honda2014perturbation}. Examples of this problem are shown in Figure~\ref{fig:expansions} where the various dashed and dotted lines represent the mean predictions from a finite-order expansion and the solid line denotes the true physical system. One can see that these models are highly accurate descriptions of the true system in some regions of the domain, yet they are unable to provide a globally accurate model. Most EFT problems fall within the
$\M$-open setting, as the true underlying description of the system across the entire domain is unknown and thus is not contained within the model set. Instead, multiple EFTs can be constructed based on the known physics to recover the true system across subsets of the domain. This poses the question as to how to combine the predictions from multiple EFTs in order to obtain a globally accurate prediction. Various interpolation methods \citep{honda2014perturbation} exist, however no data-driven approaches are currently available for EFTs.

\begin{figure}[t]
    \centering
    \includegraphics[width=0.425\textwidth, height = 0.375\textwidth]{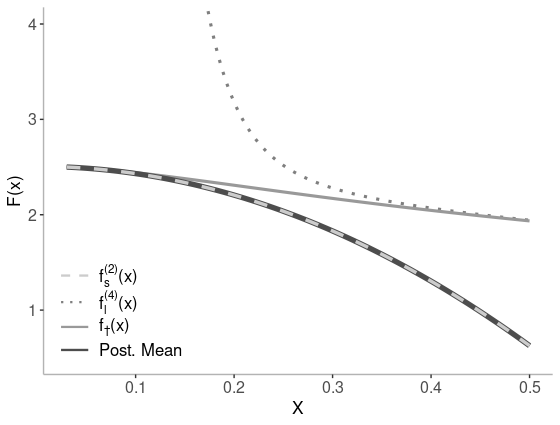}
    \caption{The posterior mean prediction of $f_\dagger(x)$ when applying BMA to the $\second$ order weak and $\fourth$ order strong coupling expansions.}
\label{fig:bma_fit}
\end{figure}

To demonstrate why problems falling in the $\M$-open class may not be suited for model averaging schemes, consider applying BMA to the model set involving the two expansions as shown in Figure~\ref{fig:expansions}(a). 
The posterior mean prediction from BMA results in a poor estimate of the true system as shown in Figure \ref{fig:bma_fit}. Essentially, BMA selects the dashed model rather than leveraging the localized strengths contained in the model set. Given the characteristics of  EFTs and the $\M$-open setting associated with these problems, a simple weighted average of the predictions from each model is insufficient for recovering the true physical system. A better approach is to use an input-dependent weighting scheme which leverages the localized behaviors of each model to ascertain appropriate mean prediction and uncertainty quantification. Such an approach falls under the general class of problems known as Bayesian model mixing (BMM) \citep{hs}.

A key challenge in BMM is to define the relationship between the inputs and the weight functions. This work proposes a Bayesian treed model which specifies the weight functions as a sum-of-trees. This representation relies on  tree bases which are used to learn the localized model behavior. Additionally, this flexible and non-parametric approach allows the user to avoid having to specify a more restrictive model for the weight functions, such as a generalized linear model. Maintaining the traditional conjugacy properties associated with Bayesian Additive Regression Tree (BART) models, the weight functions are regularized via a multivariate Gaussian prior. The prior is calibrated so that the weight functions prefer the interval $[0,1]$  without imposing any further constraints. Additionally, this framework includes a simple strategy for incorporating prior information about localized model performance when available. All together, this approach highlights the localized behaviors of the candidate models and yields significant improvements in prediction, interpretation, and uncertainty quantification compared to traditional model averaging methods.

In addition to proposing a novel non-parametric BMM method, this work introduces a new data-driven approach for combining predictions from various EFTs. This is not only important for prediction of the system, but also for the resulting inference. In particular, practitioners can better understand the accuracy of each EFT while also advancing their knowledge about the underlying physical system across areas which are not well explained by the  EFTs under consideration.  

The remainder of this paper is organized in the following manner. Section 2 highlights some relevant work related to model averaging, model mixing, and BART. Section 3 introduces the essential features of EFTs, while Section 4 outlines the specifics of the proposed BART-based framework. Three motivating EFT examples are presented in Section 5. Finally, Section 6 provides a detailed discussion of the results presented throughout this work. Full derivations of the methodology are provided in the appendix. Additional examples and information regarding EFTs are provided in the online supplement. 

\section{Background}

This section provides an overview of the primary statistical methods discussed throughout this work. Section 2.1 details popular model averaging and model mixing techniques. Section 2.2 summarizes the primary features of Bayesian tree models, which play an integral role in the proposed model mixing approach described in this work.

\subsection{Model Averaging and Model Mixing}

Methods to address model uncertainty have been widely studied throughout the past few decades. The majority of work in this area combines competing models through either mean or density estimation. In either case, the combined result is generally computed via linear combination of the individual predictive means or densities from the models under consideration. 
The weights in this linear combination may or may not depend on the inputs for each model and are learned using the set of training data $\D = \{(\xvec_1,y_1),\dots,(\xvec_n,y_n)\}$. Many frequentist and Bayesian methods exist for estimating the model weights. Popular frequentist approaches such as \textit{stacking} \citep{breiman_stacking} and \textit{model aggregation} \citep{bunea2007aggregation} estimate the weights by minimizing a specified loss function. Additionally, one may elect to impose constraints such as a non-negativity or sum-to-one constraint on the weights or apply regularization techniques. Other frequentist approaches estimate the weights using evaluation metrics such as the Akaike information criteria \citep{burnham1998practical} or Mallow's CP \citep{hansen2007least}. These methods generally fall under the model averaging regime, as the weights are independent of the model inputs, with the exception of \cite{sill2009feature}. The remainder of this section reviews popular Bayesian methods in further detail.

\noindent \textbf{Bayesian Model Averaging:} 
A classical approach for combining models $\M_1$,\ldots,$\M_K$ is  \textit{Bayesian Model Averaging} \citep{bma_lr}. Suppose $Q$ is a quantity of interest. The posterior density of Q is defined by ${\pi(Q\mid \D) = \sum_{l = 1}^K w_l \ \pi(Q \mid \D, \M_l)}$, which is a weighted average of the posterior densities with respect to each model. Each weight is defined in terms of its corresponding posterior model probability, i.e. $w_l = \pi(\M_l \mid \D)$ 
where \[\pi(\M_l \mid \D)= \frac{p(\D\mid\M_l)\pi(\M_l)}{\sum_{k=1}^K p(\D\mid\M_k)\pi(\M_k)} \] and $p(\D\mid\M_l)$ is the marginal likelihood of the data with respect to the $\lth$ model. Though BMA is useful, it has been criticized for emphasizing a fit to the training data as opposed to out-of-sample prediction, asymptotically selecting a single model (inappropriate in the $\M$-complete and $\M$-open settings, e.g. Figure~\ref{fig:bma_fit}), and being sensitive to prior specification.      

\noindent \textbf{Bayesian Mean Stacking:}
Recent work has extended \textit{stacking} to the Bayesian paradigm as an approach for mean estimation \citep{clydec2013bayesian, le2017bayes}. 
Given $K$ competing models, the stacked mean for a future observation $\tilde{y}$ at input $\tilde{\xvec}$ is constructed as a linear combination of individual model predictors 
$E[\tilde{y}\mid \tilde{\xvec},\D] = \sum_{l = 1}^K w_l \ f_l(\tilde{\xvec})$, 
where ${E[\tilde{y} \mid \tilde{\xvec},\D,\M_l] = f_l(\tilde{\xvec})}$. When the individual models are unknown, stacking is conducted in a two-step procedure: (i) independently fitting the individual models $\M_l,\ l=1,\ldots K,$ given the set of training data $\D$, and (ii) estimating the weights ${\bf w}=(w_1,\ldots,w_K)^\top$ for the stacked predictor given the fitted models. 

In the first step, each model is fit and their corresponding mean predictions, $\hat{f}_l(\xvec_i)$, are obtained at each of the training points. In practice, cross validation techniques are used to reduce the risk of overfitting the stacked predictor to the training data. In the second step, the coefficient vector $\boldsymbol{w} = (w_1,\dots,w_K)^\top$ is defined as the minimzer of a specified posterior expected loss.
Additionally, one may impose various constraints such as a simplex, non-negativity, or sum-to-m constraint on the weights \citep{le2017bayes}. Other approaches include regularization via a penalty term or a prior \citep{breiman_stacking,yang2014minimax}. 

\noindent \textbf{Bayesian Complete Stacking:}
\textit{Complete Stacking} was motivated by the shortcomings of BMA \citep{cs}. This Bayesian stacking model emphasizes prediction, as the weights are selected to minimize the Kullback-Leibler (KL) divergence between the true predictive density and the stacked predictive density $p(\tilde{y}\mid\tilde{\xvec}) = \sum_{l = 1}^K w_l \; p(\tilde{y}\mid \tilde{\xvec},\D, \M_l)$, where $\tilde{y}$ is a future observation with input $\tilde{\xvec}$. Similar to mean stacking, the leave-one-out (LOO) cross validated predictive density can be used in place of $p(\tilde{y}\mid  \tilde{\xvec},\D, \M_l)$ when the individual models are unknown. Given training data, the weights are constrained to a $K-$dimensional simplex $S_{K}$ and estimated as $\widehat{\boldsymbol{w}} = \argmax_{\boldsymbol{w} \in S_{K}} \sum_{i = 1}^n\log \sum_{l = 1}^K w_l\; p(y_i\mid \xvec_i, \D^{(-i)}, \M_l)$, where  $\D^{(-i)}$ denotes the training set excluding the pair $(\xvec_i, y_i)$. 

\noindent \textbf{Bayesian Hierarchical Stacking:}
\textit{Hierarchical Stacking} \citep{hs} is a model mixing approach which  extends Complete Stacking by defining input-dependent weights that are estimated in a fully Bayesian manner.  One way to define the weight functions is through a parametric model. First, $K-1$ unconstrained weight functions are defined as, \[w^*_l(\xvec_i) = \mu_l + \sum_{j = 1}^J\alpha_{lj}\;g_j(\xvec_i),\] which depend on the sets of hyperparameters $\{\alpha_{lj}\}$ and $\{\mu_{l}\}$ along with user-specified basis functions $g_j(\xvec_i)$, where $j = 1,\dots,J$ and $l = 1,\dots,K-1$. The $K^{th}$ function $w^{*}_K(\xvec_i)$ is set to 0 to serve as a baseline. Then, a softmax transformation is applied to the unconstrained weights in order to confine each model weight to the $K$-dimensional simplex, namely \[w_l(\xvec_i) = \frac{\exp\big(w^*_l(\xvec_i)\big)}{\exp\big(w^*_1(\xvec_i)\big) +\dots+\exp\big(w^*_K(\xvec_i)\big)}, \quad l = 1,\dots,K.\]  

The methods discussed above outline a number of strategies one can take to combine the information across multiple models. In the setting of EFT experiments, the localized nature of the predictions suggests an input-dependent weighing scheme like Bayesian Hierarchical Stacking is more suitable.  However, specifying the required basis functions may not be trivial. Thus, the proposed method will adopt the notion of mean stacking within an additive tree basis model to achieve localized weighting in a flexible and non-parametric manner.  

\subsection{Bayesian Tree Models}
Bayesian additive regression trees (BART) have become increasingly popular for modeling complex and high dimensional systems \citep{bart_2010}. 
This additive approach involves summing together the predictions made from $m$ trees and is facilitated through a Bayesian backfitting algorithm \citep{hastie2000bayesian}. Each tree $T_j$ is characterized by its structure, comprised of internal and terminal nodes, along with its associated set of terminal node parameters, $M_j$. The internal nodes define binary partitions of the input space according to a specified splitting rule. 
A given node $\eta$ is defined to be an internal node with probability $p(\eta \text{ is internal}) = \alpha(1 + d_\eta)^{-\beta}$ where $d_\eta$ is the depth of $\eta$ and $\alpha$ and $\beta$ are tuning parameters. By construction, this prior penalizes tree complexity and thus ensures each tree maintains a shallow and simple structure. Given $d$ different predictors, $x_1,...,x_d$, splitting rules are of the form $x_v < c$ for $v \in \{1,...,d\}$ and cutpoint $c$ from a discretized subset of $\R$. 
In the simplest approach, the predictor and cutpoint associated with each splitting rule are randomly selected from discrete uniform distributions. The probabilities associated with the designation of each node along with the splitting rules for internal nodes are used to define the stochastic tree-generating prior for each tree. 

The $m$ trees are learned through MCMC, where a slight modification to each structure is proposed at every iteration of the simulation. Generally, such modifications to the tree include birth, death, perturb, or rotate as described by \cite{bcart} and \cite{pratola2016efficient}. Proposals are then accepted or rejected using a Metropolis-Hastings step. To avoid a complex reversible jump MCMC, the algorithm depends on the integrated likelihood, which is obtained by integrating over the terminal node parameters associated with the given tree. A closed form expression for this density can be obtained with conditional conjugate priors for the terminal node parameters.  

Given the tree structure, prior distributions can be assigned to each terminal node parameter. 
In the BART model, the priors ensure each tree explains a small yet different source of variation in the data. For continuous data, BART assigns Gaussian priors to the terminal node parameters. Assuming the data is mean centered, the prior assigned to terminal node parameter $\mu_{pj}$ in node $\eta_{pj}$ is given by $\mu_{pj}\mid T_j\sim N(0,\tau^2)$ where $\tau = (y_{max} - y_{min})/(2k\sqrt{m})$ with tuning parameter $k$.  Additionally, a scaled inverse chi-squared prior is assigned to the variance, i.e. $\sigma^2\sim \nu\lambda/\chi^2_\nu$. 

The traditional Bayesian regression tree model can be extended to allow for a more complex structure in the terminal nodes.  Existing extensions include linear regression 
\citep{bayes_treed,prado2021bayesian} 
and Gaussian processes \citep{bayes_tree_gp}.  For the setting of model mixing, this work utilizes a multivariate Gaussian terminal node model.

\section{Towards Model Mixing with EFTs}

An EFT forms an expansion (or multiple expansions) as a ratio of an input parameter to a physically relevant scale. Computer models implement EFTs as simulators. The theoretical predictions of the physical system are approximations from each simulator plus a discrepancy term, which is designed to account for the remaining unexplained portions of a system. These two components may have specific properties which can be leveraged when working with observational data. This section summarizes these details in the context of EFTs, while a further discussion is provided in the supplementary material. 

\subsection{Motivating EFT Example} \label{honda_eft_section}
Consider the EFT example where the true physical system is the partition function of the zero-dimensional $\phi^4$ theory defined by \begin{equation} \label{true_phy}
    f_\dagger(x) = \int_{-\infty}^{\infty} \exp\Big(-\frac{u^2}{2} - x^2u^4\Big)\;du,
\end{equation} 
where $x$ denotes the coupling constant \citep{honda2014perturbation}.  Two types of finite-order expansions exist for this partition function and are given by (\ref{h_sg}) and (\ref{h_lg}) for $n_s$ or $n_l$ $ \geq 1$, namely 
\begin{align}
    h_s^{(n_s)}(x) &= \sum_{t = 0}^{n_s}s_t x^t \quad \text{where} \;\; 
    s_{t} = \begin{cases} \frac{\sqrt{2}\Gamma(t + 0.5)}{(t/2)!}(-4)^{(t/2)} & t \text{ is even} \\[4 pt] 0 & t \text{ is odd}
    \end{cases} \label{h_sg} \\[5 pt]
    h_l^{(n_l)}(x) &= \sum_{t = 0}^{n_l}l_t x^{-t} \quad \text{where} \;\; l_t = \frac{\Gamma(0.5t+ 0.25)}{2t!}\Big(-\frac{1}{2}\Big)^t, \quad t = 0,...,n_{l}. \label{h_lg}
\end{align}


The weak coupling expansion in (\ref{h_sg}) is an asymptotic Taylor-like series of order $n_s$ centered about zero. Thus, $h_s^{(n_s)}(x)$ will yield high-fidelity predictions for smaller coupling constants and diverge as the value increases. The reverse behavior is observed for the strong coupling expansion in (\ref{h_lg}),  $h_l^{(n_l)}(x)$, which is convergent. 
Example predictions of the physical system using these finite-order expansions can be seen in Figure~\ref{fig:expansions} and are discussed in detail in Section \ref{model_set}. 

The theoretical predictions of the physical system using the weak and strong coupling expansions are expressed using (\ref{fs_eft}) and (\ref{fl_eft}), respectively. 
\begin{align}
    f_s^{(n_s)}(x) &= h_s^{(n_s)}(x) + \delta_s^{(n_s)}(x) \label{fs_eft} \\[5 pt] f_l^{(n_l)}(x) &= h_l^{(n_l)}(x) + \delta_l^{(n_l)}(x). \label{fl_eft}
\end{align}

\noindent where the truncation errors $\delta_s^{(n_s)}(x)$ and $\delta_l^{(n_l)}(x)$ are modeled with Gaussian processes (GPs) ~\citep{gramacy2020surrogates,santner2018design}. As described by \cite{CTE_EFT}, the parameters in both truncation error models are dependent upon the evaluations of their corresponding finite-order expansions (described in (\ref{h_sg}) and (\ref{h_lg}), respectively) over a sparse grid of points. The discrepancy model also depends on physical quantities, $Q(x)$ and $y_{\text{ref}}(x)$, which are chosen based on domain expertise. The relationship between these quantities and the discrepancy are summarized in the supplementary material. When $Q(x)$ and $y_{\text{ref}}(x)$ are unknown, one can alternatively use the error approximation described by \cite{semposki2022uncertainties}.



The features present in this example from \cite{honda2014perturbation} are commonly found across the landscape of EFT problems. For instance, the physical system can be expressed as an additive model involving a finite-order expansion and the induced truncation error. The finite-order expansions are designed to provide high-fidelity predictions in specific subregions of the domain. There exists a subregion of the domain where none of the finite-order expansions yield accurate theoretical predictions. 
All together, this motivating example serves as a prototype for the EFTs that may be encountered in a general experimental setting.  

\subsection{The Model Set for EFT Experiments} \label{model_set}

One may encounter various experimental settings when working with EFTs. Such scenarios are introduced in the context of the motivating
example presented in Section \ref{honda_eft_section}. First, consider the most basic case where the model set contains a single EFT. 
With one EFT, the overall predictive accuracy of the true system is poor, despite the good performance in a localized region. For example, suppose the model set $\M$ contains the $\second$ order weak coupling expansion $f_s^{(2)}(x)$. Mean predictions constructed from (\ref{h_sg}) and (\ref{fs_eft}) are shown by the dashed line in Figure~\ref{fig:expansions}(a). Clearly, this model is  limited to strong predictive accuracy in only the left subregion of the domain. 

When available, one can consider different finite-order approximations of the same EFT. For example, consider the $\second$, $\fourth$, and the $\sixth$ order coupling expansions which are shown in Figure~\ref{fig:expansions}(c). The three models are very similar for lower coupling constants yet drastically differ in the remainder of the domain. Despite each expansion's poor theoretical predictions, one can still leverage the available information to improve the overall prediction of the physical system. For instance, the $\second$ and $\sixth$ order expansions (dashed and dashed-dotted) are concave functions while the $\fourth$ order expansion (dotted) is convex. This suggests the true physical system lies between the expansions under consideration and can be recovered by re-weighting the corresponding predictions. 

A third situation is to consider EFTs centered about different areas of the domain. For example, a model set can contain a finite-order weak coupling expansion (dashed) and the $\fourth$ order strong coupling expansion (dotted) as shown in Figures \ref{fig:expansions}(a) and \ref{fig:expansions}(b). The addition of the strong coupling expansion allows for a high-fidelity approximation of the physical system to be considered in the rightmost subregion of the domain. The model set listed in panel (a) implies the true system lies between the two expansions. This is particularly useful in the intermediate range where neither of the EFTs are accurate. Meanwhile, the set in panel (b) presents an interesting case where the physical system lies below both EFTs in the intermediate range. In this case, the information in the observational data can be leveraged to help recover the true system. 

In this example, the predictions from the weak coupling expansion  degrade slowly compared to those from the strong coupling expansions. Consequently, the weak coupling expansions generally appear to have a better overall predictive performance across the entirety of the domain. When combining these two types of EFTs using global weighting schemes such as BMA, the resulting prediction will favor the weak coupling expansion due to its drastic advantage in the overall model performance. The undesirability of the BMA solution is evident in Figure \ref{fig:bma_fit}, which demonstrates that BMA effectively matches the $\second$ order weak coupling expansion. Hence, a weighting scheme which captures the localized behaviors of each model is preferred in the EFT setting.     

The proceeding sections consider a general set of $K$ different EFTs, which are denoted by $f_1(x),\ldots,f_K(x)$. In this motivating example, $f_l(x) = h_l(x) + \delta_l(x)$ where $h_l(x)$ can denote either a weak or strong coupling expansion of order $N_l$, where $l = 1,\ldots,K$. Meanwhile, $\delta_l(x)$ is the associated truncation error and is modeled by a GP as described in the supplementary material.    

\subsection{Predictions from EFTs} \label{eft_predictions}

Prior to model mixing, each of the K EFTs are independently emulated.
Without loss of generality, consider the $\lth$ EFT denoted by $f_l(\xvec)$. It is assumed this EFT is accompanied by a set of simulator runs across a fixed set of inputs $\xvec^c_{l1},\ldots,\xvec^c_{ln_l}$. Information regarding the design of the computer experiment for each EFT can be found in ~\cite{melendez2021designing}. The simulator runs are evaluations of the finite-order expansion, $h_l(\cdot)$, at the specified inputs. Using these runs, one can extract the set of $N_l+1$ coefficients $c_0(\cdot),\ldots,c_{N_l}(\cdot)$ at each of the fixed inputs. The training set for the $\lth$ EFT is then defined by  ${\D_l = \big\{\big(\xvec^{c}_{l1},\boldsymbol{C}(\xvec^{c}_{l1})\big),\ldots,(\xvec^{c}_{ln_l},\boldsymbol{C}(\xvec^{c}_{ln_l})\big)\big\}}$ where $\boldsymbol{C}(\cdot)$ denotes the vector of known finite-order coefficients at the specified model input. The resulting coefficients and the set of inputs can differ across the $K$ models, thus the sets $\D_1,\ldots,\D_K$ will contain different information. 

As described in the supplementary material, an EFT is fit using the finite-order coefficients to learn the unknown parameters which characterize the GP assigned to the truncation error. This information can be extracted from $\D_l$, which implies the set of field observations is not required to fit each EFT. Consequently, the desired theoretical predictions across the input domain can be obtained without using any of the observational data. 
The resulting posterior predictive distribution is a Gaussian process, which can be characterized by the corresponding mean and covariance functions as described in \cite{CTE_EFT} (see also the supplementary material).
The predictions for an EFT are then computed through the posterior mean. 


\section{Bayesian Additive Model Mixing Trees}

\subsection{Defining a Mixed Model}
The proposed BMM model is trained using a set of observational data, $Y_1,...,Y_n$, which are assumed to be independently generated at fixed inputs $\xvec_1,...,\xvec_n$ according to 
\begin{equation*} Y_i = f_\dagger(\xvec_i) + \epsilon_i, \quad  \epsilon_i \iid N(0,\sigma^2)  \end{equation*} 
where  $f_\dagger(\xvec_i)$ represents the true and unknown physical system. Conditional on the theoretical predictions at a given point, $f_1(\xvec_i),\ldots,f_K(\xvec_i)$, the data can be modeled as 
\begin{equation}\label{mix_model1}
    Y_i \mid \fvec(\xvec_i), \wvec(\xvec_i), \sigma^2 \ind N\big(\fvec^\top(\xvec_i) \wvec(\xvec_i),\sigma^2\big)
\end{equation}

\noindent where $\fvec(\xvec_i) = \big(f_1(\xvec_i),...,f_K(\xvec_i) \big)^\top$ and $\wvec(\xvec_i) = (w_1(\xvec_i),...,w_K(\xvec_i))^\top$. This formulation is an example of Bayesian mean stacking with an input-dependent weighting scheme. In practice, the predictions from each model are unknown and must be estimated. 

The proposed BMM model relies on a two-step approach for combining the predictions across $K$ EFTs. 
This implies each EFT is first fit independently and the estimated predictions $\hat{f}_l(\xvec_i)$ are obtained for $l=1\ldots,K$ and $i=1,\ldots,n$ prior to learning the weight functions $w_1(\xvec_i),\ldots,w_K(\xvec_i)$. The proposed two-step approach is tailored to EFTs by taking advantage of the sources of data described in Section \ref{eft_predictions} as well as the properties described in the supplementary material. Conditional on the estimated predictions, the model for the observational data becomes 
\begin{equation*}\label{mix_model2}
    Y_i \mid \fhatvec(\xvec_i), \wvec(\xvec_i), \sigma^2 \ind N\big(\fhatvec^\top(
    \xvec_i) \wvec(\xvec_i),\sigma^2)
\end{equation*}

\noindent where $\fhatvec(\xvec_i) = \big(\hat{f}_1(\xvec_i),\ldots,\hat{f}_K(\xvec_i) \big)^\top$. The weight functions are then learned using the set of field data. The next section outlines the proposed model mixing scheme which defines the weight functions using Bayesian Additive Regression Trees (BART).


\subsection{Model Mixing using BART} \label{MixBART}

The weight functions ${\bf w}(\xvec)=\left(w_1(\xvec),\ldots,w_K(\xvec)\right)^\top$ are modeled using a sum-of-trees 
\begin{equation} \label{wts_sum_of_trees}
    \wvec(\xvec_i) = \sum_{j=1}^m \gvec(\xvec_i,T_j,M_j),    
\end{equation}
\noindent where $\gvec(\xvec_i,T_j,M_j)$ is the $K$-dimensional output of the $\jth$ tree using the set of terminal node parameters, $M_j$, at the input, $\xvec_i$. 
This approach defines the weight functions using tree bases which are learned from the data. The amount of flexibility in the weight functions can be controlled by changing the number of trees or tuning the hyperparameters in the prior distributions. 

In this application of BART, each terminal node parameter is a $K$-dimensional vector which is assigned a multivariate Gaussian prior. The parameter is regularized 
so that each tree accounts for a small amount of variation in the weight functions. For the proceeding statements, let $\eta_{pj}$ represent the $\pth$ terminal on the $\jth$ tree and define its corresponding  parameter by $\muvec_{pj} = (\mu_{pj1},...,\mu_{pjK})^\top$. Now assume the observations $(\xvec_1,y_1),...,(\xvec_{n_p},y_{n_p})$ lie in the hyper-rectangle defined by $\eta_{pj}$, where $n_p$ is the number of observations assigned to this subregion. The model at each terminal node amounts to fitting a localized Bayesian linear regression with parameter vector $\muvec_{pj}$. Due to conditional independence, the likelihood in this node is defined by    
\begin{equation*}
    L(r_1,...,r_{n_p} \mid T_j, \muvec_{pj}, \sigma^2) = (2\pi\sigma^2)^{-n_p/2}\exp\Big(-\frac{1}{2\sigma^2} \sum_{i = 1}^{n_p}\Big(r_i - \fhatvec^\top(\xvec_i)\muvec_{pj}\Big)^2\Big)
\end{equation*}

\noindent where $\fhatvec(\xvec_i) = \big(\hat{f}_1(\xvec_i),...,\hat{f}_K(\xvec_i)\big)^\top$ is a vector of mean predictions from each EFT and $r_i$ is the $\ith$ residual given by $r_i = y_i - \sum_{q\ne j}\fhatvec^\top(\xvec_i)g(\xvec_i, T_q, M_q).$

Conditional on the tree structure, $T_j$, the terminal node parameter, $\muvec_{pj}$ is assigned a conjugate multivariate Gaussian prior, namely 
\begin{equation} \label{noninform_prior} \muvec_{pj} \mid T_j \ind N_K\Big(\betavec, \tau^2 \boldsymbol{I}_K \Big)\end{equation} where $\betavec = (\beta_{1},...,\beta_{K})^\top$ is a $K$-dimensional mean vector and $\boldsymbol{I}_K$ is the identity matrix. This prior is non-informative in the sense that the mean is fixed regardless of how the input space is partitioned. 


In model mixing, each simulator may perform strongly in one subregion of the input space but weakly in another. This belief can be reflected in the prior distribution of $\muvec_{pj}$ by allowing the hyperparameters to depend on the partition of input space assigned to the given terminal node. Thus, an informative prior for $\muvec_{pj}$ can be constructed as  
\[ \muvec_{pj} \mid T_j \ind N_K\Big(\betavec_{pj}, \tau^2 \boldsymbol{I}_K \Big)\]
\noindent where $\betavec_{pj} = (\beta_{pj1},...,\beta_{pjK})^\top$. This allows the prior mean to vary depending on the tree partitions and thus reflect some sense of localized model performance. Meanwhile, the assumed covariance structure implies the $K$ vector components $\mu_{pj1},...,\mu_{pjK}$ are independent apriori. 

Both of the proposed priors are conjugate, which is an important choice in BART, as it allows for a closed form expression for the marginal likelihood for the vector of residuals $\rvec_{pj} = (r_1,..,r_{n_p})^\top,$
%
%
Additionally, the conjugate priors result in closed form expressions for the full conditional distributions of the terminal node parameters and the error variance. The derivations of these distributions are found in the Appendix.  In particular, the full conditional  distribution for the $\pth$ terminal node in $T_j$ is given by 
\[ \muvec_{pj} \mid \rvec_{pj}, T_j, \sigma^2 \ind N_K\Bigg(\Big(\frac{1}{\sigma^2}\hat{\Fvec}^\top_{pj}\hat{\Fvec}_{pj} + \frac{1}{\tau^2}\boldsymbol{I}_K \Big)^{-1}\Big(\frac{1}{\tau^2}\betavec_{pj} + \frac{1}{\sigma^2}\hat{\Fvec}^\top_{pj}\rvec_{pj}\Big), \Big(\frac{1}{\sigma^2}\hat{\Fvec}^\top_{pj}\hat{\Fvec}_{pj} + \frac{1}{\tau^2}\boldsymbol{I}_K \Big)^{-1}\Bigg)\]
where $\hat{\Fvec}_{pj}$ is the $n_p \times K$ design matrix with the $\ith$ row vector given by the vector $\fhatvec^\top(\xvec_i)$. The full conditional  distribution for $\sigma^2$ is a scaled inverse chi-squared, i.e. $\sigma^2 \mid \cdot \sim \nu^\prime\lambda^\prime/\chi^2_{\nu^\prime}$, where \[\nu^\prime = n+\nu \quad \text{and} \quad \lambda^\prime = \frac{1}{n+\nu}\Big(\sum_{i = 1}^n \Big(y_i - \fhatvec^\top(\xvec_i)\wvec(\xvec_i)\Big)^2 + \nu \lambda \Big),\]  
 with $\nu$ and $\lambda$ denoting the prior shape and scale parameters, respectively.

\subsection{Calibrating Priors}

First consider the prior for the terminal node parameters. The calibration of the hyperparameters differs for the non-informative and informative priors, however both approaches are designed to ensure that each model weight $w_l(\xvec)$ should prefer the interval $[0,1]$ and be centered at a value within this region. Moreover, the functions $w_1(\xvec),\ldots,w_K(\xvec)$ are assumed to be  independent apriori at a fixed input. This enables the prior for each weight to be calibrated marginally.


\subsubsection{Non-Informative Prior}
Consider a non-informative prior for the terminal node parameters. In this setting,\\ ${\muvec_{pj} \mid T_j \iid N_K\big(\betavec, \tau^2 \boldsymbol{I}_K \big)}$ for the $\pth$ terminal node parameter in the $\jth$ tree. First, fix $l \in \{1,...,K\}$ and $i \in \{1,...,n\}$ to calibrate the prior for $w_l(\xvec_i)$. Since the terminal node parameters are independent and identically distributed with a diagonal covariance structure, the prior induced on $w_l(\xvec_i)$ is the same for the remaining weight and input combinations. From (\ref{wts_sum_of_trees}) and (\ref{noninform_prior}), the induced prior on the $\lth$ model weight is $w_l(\xvec_i) \sim N(m\beta_l, m\tau^2)$. Since it is believed $w_l(\xvec_i) \in [0,1]$ with high probability, it is plausible to set $m\beta_l =0.5$. Consequently, $\beta_l =0.5/m$. Thus, each weight has an equal chance to reach the ``extreme" values of 0 or 1 regardless of the input location. The prior standard deviation, $\tau$, can be  selected  so that $w_l(\xvec_i) \in [0,1]$ with high probability. To do this, a confidence interval for $w_l(\xvec_i)$ is constructed such that $0 = 0.5 - k\tau\sqrt{m}$ and 
$1 = 0.5 + k\tau\sqrt{m}$.  Subtracting the first equation from the second and solving for $\tau$ yields $\tau = 1/2k\sqrt{m}$. This calibration approach is very similar to the one proposed by \cite{bart_2010}. The main difference is due to the context of the problem, as it is believed the weights are predominately contained in an interval $[0,1]$ rather than the observed range of the data, $[y_{min}, y_{max}]$.

\subsubsection{Informative Prior}

In the informative setting, the prior mean directly depends on the partitions of the input space induced by the given tree, i.e. $\muvec_{pj} \mid T_j \sim N_K\big(\betavec_{pj}, \tau^2 \boldsymbol{I}_K\big)$. This prior is tailored towards EFTs, where the functional variance, $v_l(\xvec_i)$, indicates the severity of the truncation error. A larger variance within a particular subregion of the domain indicates the presence of larger truncation error meaning the EFT provides a poor approximation of the true system.

Given this interpretation of the truncation error variances, one strategy for combining EFTs is precision weighting \citep{band2021}. For example, the precision weight for the $\lth$ EFT at $\xvec_i$ is given by   
\[\beta_l(\xvec_i) = \frac{1/v_l(\xvec_i)}{1/v_1(\xvec_i) + ... + 1/v_K(\xvec_i)}. \]
The precision weight $\beta_l(\xvec_i)$ can be interpreted as an initial guess for the weight function $w_l(\xvec_i)$ for $l=1,\ldots,K$ and $i=1,\ldots,n$. 

Since the prior of the terminal node parameter changes conditional on the tree structure, each $\betavec_{pj}$ is chosen separately from the other terminal node parameters. Given the precision weights for the EFTs and $n$ training points, each component of the the prior mean vector, $\beta_{pjl}$, is chosen by  
\[\beta_{pjl} = \frac{1}{m \sum_{i=1}^n \onevec(\xvec_i\in\eta_{pj})} \sum_{i = 1}^{n} \beta_l(\xvec_i) \;\onevec(\xvec_i\in\eta_{pj}),\]
where $\onevec(\xvec_i \in \eta_{pj})$ is the indicator that $\xvec_i$ is assigned to the terminal node $\eta_{pj}$.
A confidence interval for each terminal node parameter can be set to have a length of $1/m$ in order to ensure each tree is a weak learner. This is done by setting $\tau = 1/2km$. 
    
\subsubsection{Variance Prior}

A conjugate scaled inverse chi-squared distribution with hyperparameters $\nu$ and $\lambda$ is assigned to the error variance $\sigma^2$.
To calibrate the prior, first select a value of $\nu$ to reflect the desired shape of the distribution. Common values of $\nu$ range from 3 to 10. Before selecting a value for $\lambda$, one needs an initial estimate of the error variance to help set the prior around a range of plausible values of $\sigma^2$. Given the model set and the corresponding point predictions at each of the training points $\boldsymbol{\hat{f}}_l(\xvec_i)$, one can use a lightly data informed prior by setting $\hat{\sigma}^2 = \max_{l = 1,..,K} \Big\{\min_{i=1,..,n} \Big(y_i - \hat{f}_l(\xvec_i)\Big)^2  \Big\}$. Since a common belief is that each model yields accurate approximations of the true system over some subregion of the domain, one should expect the set of minimum squared differences across the $K$ models will unveil reliable information about the true error variance. Given this information, one strategy is to set $\hat{\sigma}^2$ to be the mean or mode of a $\lambda\nu/\chi_\nu^2$ distribution. The value of $\lambda$ is then found by solving the resulting equation.  


\section{EFT Examples}

This section applies the proposed model mixing methodology to three different examples. Section \ref{eft_ex_section} demonstrates the success of the BART-based mixing approach on two univariate EFT examples, which are introduced in Section 3. A multi-dimensional example is highlighted in Section \ref{ex2_2d} using simulators which are based on Taylor series expansions of a trigonometric function. Though this last example does not involve a true underlying physical system, the model set considers simulators which have similar qualities of EFTs with double expansions (see \cite{burgess_2020}). Each example highlights specific features of the proposed BART-based mixing model such as flexible basis functions for the weights and the associated prior regularization. 



\subsection{Example 1: Mixing Univariate EFTs} \label{eft_ex_section}

This section applies the BART model mixing (BART-BMM) method to various EFTs over a one-dimensional domain. For comparison, Hierarchical Stacking (HS) is also applied to the same set of EFTs. In both EFT examples, 20 observations are independently generated according to \[Y_i = f_\dagger(x_i) + \epsilon_i, \quad \epsilon_i \sim N(0,\sigma^2)\] 
where $i = 1,...,20$, $\sigma = 0.005$, and $f_\dagger(x)$ is defined in (\ref{true_phy}). The 20 training points are located at inputs which are evenly spaced over the interval of 0.03 to 0.50. The error standard deviation of $0.005$ was selected to mimic a controlled experiment setting. Each EFT model is fit using $n_c = 4$ evaluations of the corresponding finite-order expansion.

\subsubsection{Example 1a: Mixing Two EFTs} \label{ex1a}
\begin{figure}[h]
    \centering
    \includegraphics[width = 0.95\textwidth, height = 0.70\textwidth]{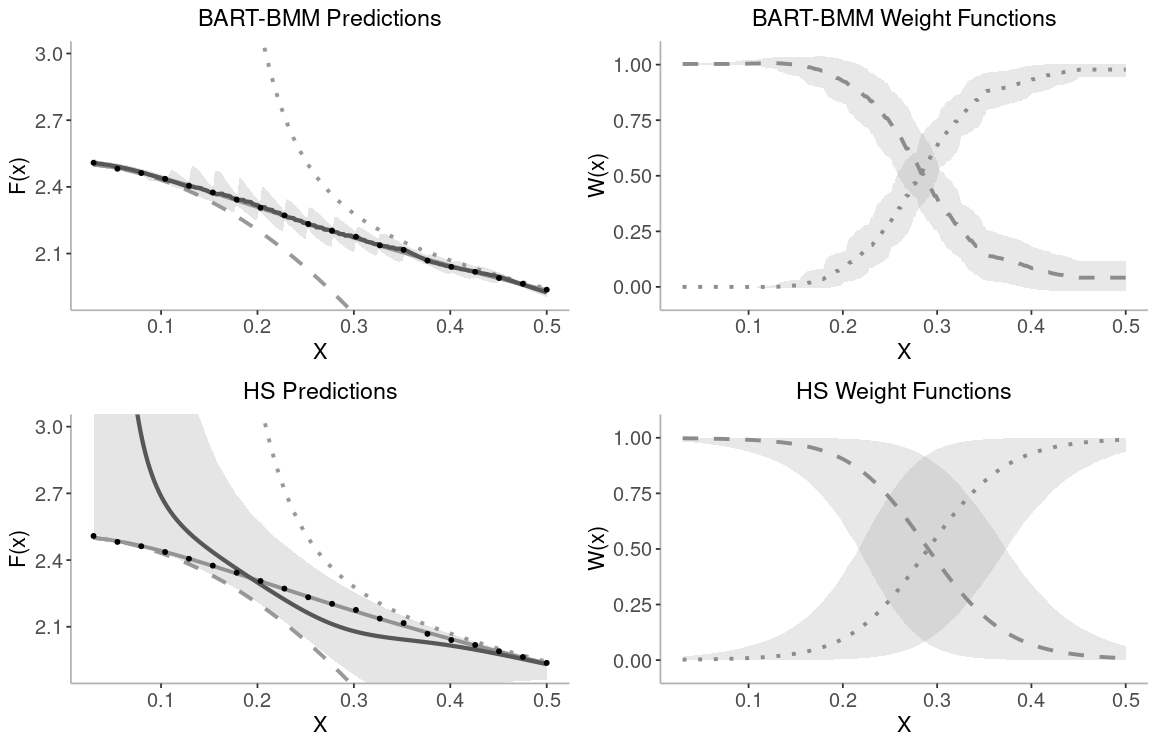}
    \caption{The predicted mean (dark gray) and 95\% credible intervals (shaded) when mixing $f^{(2)}_s(x)$ (dashed) and $f^{(4)}_l(x)$ (dotted). Results are obtained from a BART-BMM model with 10 trees and a Hierarchical Stacking model with a linear unconstrained weight function (bottom).}
    \label{fig:ex1_sg2lg4}
\end{figure}

First consider mixing the EFTs based on the second order weak coupling expansion, $f_s^{(2)}(x)$ and the fourth order strong coupling expansion, $f_l^{(4)}(x)$ as shown in Figure~\ref{fig:expansions}(a). The true system $f_\dagger(x)$ lies between both EFTs across the entire domain, hence a convex combination of the predictions from both EFTs is appropriate for recovering the true system. The BART-BMM model is fit using 10 trees and $k=5.0$. Meanwhile, the HS unconstrained weight function is defined by $w_1^{*}(x) = \mu_1 + \alpha_1x$. The results of the BART-BMM method and HS are shown in Figure \ref{fig:ex1_sg2lg4}. 

In terms of the root mean squared error (RMSE) between the predicted system and the true $f_\dagger(x)$, the BART-BMM model results in more accurate mean predictions compared to HS, which have RMSE values of 0.0053 and 1.9460 respectively. The RMSE for the HS result is inflated by the diverging mixed prediction in the left portion of the domain. For example, the RMSE for the HS model over the interval $[0.1,0.5]$ drops to 0.0717. Additionally, from Figure \ref{fig:ex1_sg2lg4} it is evident BART-BMM results in predictions of $f_\dagger(x)$ which have lower uncertainty than those from HS. 

The weight functions in Figure \ref{fig:ex1_sg2lg4} also take similar sigmoid-like shapes, however the HS solution displays a high degree of uncertainty. The most noticeable difference between the two methods can be seen in the weight function of $f^{(4)}_l(x)$ (dotted). In particular, the curve in the BART-BMM result increases at a quicker rate in the sub-region $[0.3,0.4]$ compared to the HS result. This slower rate of increase contributes to the poor prediction from HS in this sub-region. Another difference is observed in the region of $[0.03, 0.15]$, as the weight of $f^{(4)}_l(x)$ under the HS approach is near 0, however it is not small enough to negate the effect of the drastically diverging mean prediction from $f^{(4)}_l(x)$. Meanwhile, the BART-BMM weight is shrunk close to 0 with minimal uncertainty due to the mean estimation objective, which directly re-weights the mean prediction from an individual model, and the lack of a simplex constraint.       


Another advantage of BART-BMM is that the weight functions are learned throughout the MCMC via the tree models. This differs from HS, which requires specification of a basis for the unconstrained weights apriori. In this example, one may consider a different basis function, as the specified linear basis appears to be inadequate for ascertaining high-fidelity mean predictions across the entire domain. 

\subsubsection{Example 1b: Mixing Two Convex EFTs} \label{ex1b}

Now, consider a second model set which is shown in Figure~\ref{fig:expansions}(b) and replaces $f_s^{(2)}(x)$ with $f_s^{(4)}(x)$. Both EFTs overestimate $f_\dagger(x)$ in the intermediate range, hence weights which are confined to a simplex are unable to recover the true system. In this case, a piecewise basis function is assigned to the unconstrained HS weight as shown below, 
\[w_1^{*}(x) = \mu_1 + \alpha_1 \onevec(x<0.15) + \alpha_2 \onevec(0.15\leq x<0.25) + \alpha_3 \onevec(0.25\leq x<0.35). \]
This basis was chosen to roughly reflect the areas where the mean predictions begin to change at differing rates. Other selections of the partitions for a piecewise basis are equally valid.

\begin{figure}[t]
    \centering
    \includegraphics[width = 0.95\textwidth, height = 0.70\textwidth]{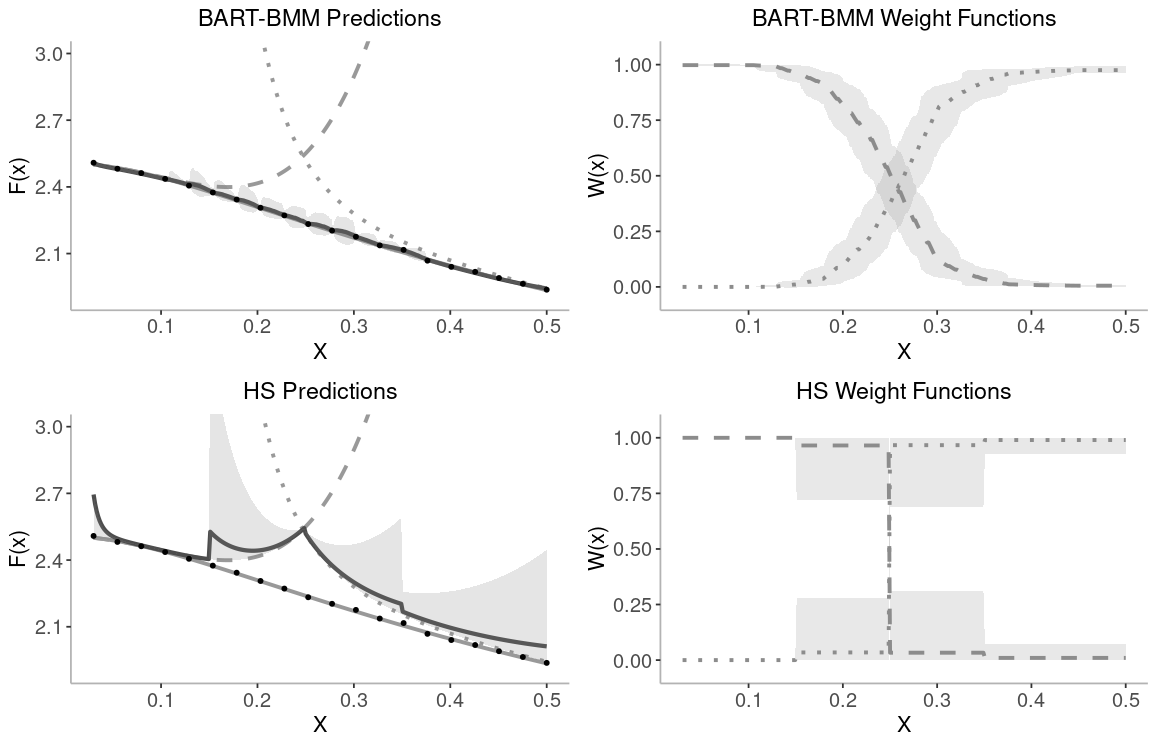}
    \caption{The predicted mean (dark gray) and 95\% credible intervals (shaded) when mixing $f^{(4)}_s(x)$ (dashed) and $f^{(4)}_l(x)$ (dotted). Results are obtained from a BART-BMM model with 10 trees (top) and a Hierarchical Stacking model with a piecewise unconstrained weight function (bottom).}
    \label{fig:ex1_sg4lg4}
\end{figure}

The BART-BMM and HS results are shown in Figure \ref{fig:ex1_sg4lg4}. Once more, the BART-BMM approach outperforms HS in terms of mean prediction, with RMSE values of 0.0057 and 0.1141 respectively. Most notably, the HS solution is unable to accurately predict the true system in the intermediate range of the domain due to the simplex constraint on the model weights. Meanwhile, the BART-BMM approach is able to recover the system across the entirety of the domain due to the prior regularization approach taken with the weights, which does not impose such strict constraints.  

In this HS result, it appears the piecewise basis was more effective than the linear basis in terms of predicting the true system in the left and right portion of the domain. This further poses the question of how to select the partitions induced by the piecewise basis, as different choices may lead to drastically different results. This question served as the motivation for defining a BART-based model, which adaptively learns these partitions based on the observational data and the model set.  

\subsection{Example 2: Multi-Dimensional Mixing} \label{ex2_2d}

The proposed model mixing approach is also applicable for computer experiments which depend on multi-dimensional inputs. To demonstrate this, consider a 2-dimensional problem where the true underlying system is defined by \[f_\dagger(\xvec) = \sin(x_1) + \cos(x_2),\]
where $\xvec =(x_1,x_2)^\top \in [-\pi,\pi]\times[-\pi,\pi].$ A set of 80 training points are generated from this true system with observational error standard deviation of $0.1$. Additionally two candidate models are considered, each with simulators defined in terms of Taylor series expansions of $s(x_i) := \sin(x_1)$ and $c(x_2):=\cos(x_2)$. For this example, the simulators are defined by 

\begin{align*}
    h_1(\xvec) = \sum_{j=0}^7 \frac{s^{(j)}(x_1)}{j!}(x_1-\pi)^j + \sum_{k=0}^{10} \frac{c^{(k)}(x_2)}{k!}(x_2-\pi)^k  \\[5 pt]  
    h_2(\xvec) = \sum_{j=0}^{13} \frac{s^{(j)}(x_1)}{j!}(x_1+\pi)^j + \sum_{k=0}^6 \frac{c^{(k)}(x_2)}{k!}(x_2+\pi)^k
\end{align*}
where $s^{(j)}(x_1)$ and $c^{(k)}(x_2)$ denote the $\jth$ and $\kth$ derivatives of $\sin(x_1)$ and $\cos(x_2)$, respectively. Note, the first simulator $h_1(\xvec)$ centers both Taylor series expansions about $\pi$, hence it produces relatively accurate predictions of the system in upper right corner of the domain and diverges when moving towards the negative portion of the domain. Meanwhile, the $h_2(\xvec)$ is composed of Taylor series  expansions centered about $-\pi$ which produces accurate predictions in the negative portion of the domain. One key difference between the simulators is that $h_2(\xvec)$ contains a highly accurate approximation of $\sin(x_1)$ across the entire interval $[-\pi,\pi]$ because its corresponding Taylor series expansion is composed of 7 non-zero terms. Thus, even though the expansion $\sin(x_1)$ and $\cos(x_2)$ are centered about $-\pi$, one would expect $h_2(\xvec)$ to result in accurate predictions of $f_\dagger(\xvec)$ across the rectangle $[-\pi,\pi]\times[-\pi,0]$.

The theoretical predictions from each model $f_1(\xvec)$ and $f_2(\xvec)$ can be defined using the additive form $f_l(\xvec) = h_l(\xvec) + \delta_l(\xvec)$, where $\delta_l(\xvec)$ represents the unknown higher-order corrections and $l=1,2$. Due to the nature of this example, no model is postulated for $\delta_l(\xvec)$. Consequently, the estimated theoretical predictions at each training point $\xvec_i$ are obtained by $\hat{f}_l(\xvec_i) = h_l(\xvec_i)$. Note, in a multi-dimensional EFT setting, the strategy discussed from Section 3 remains applicable.  

\begin{figure}[t]
    \centering
    \includegraphics[width = 1\textwidth, height = 0.50\textwidth]{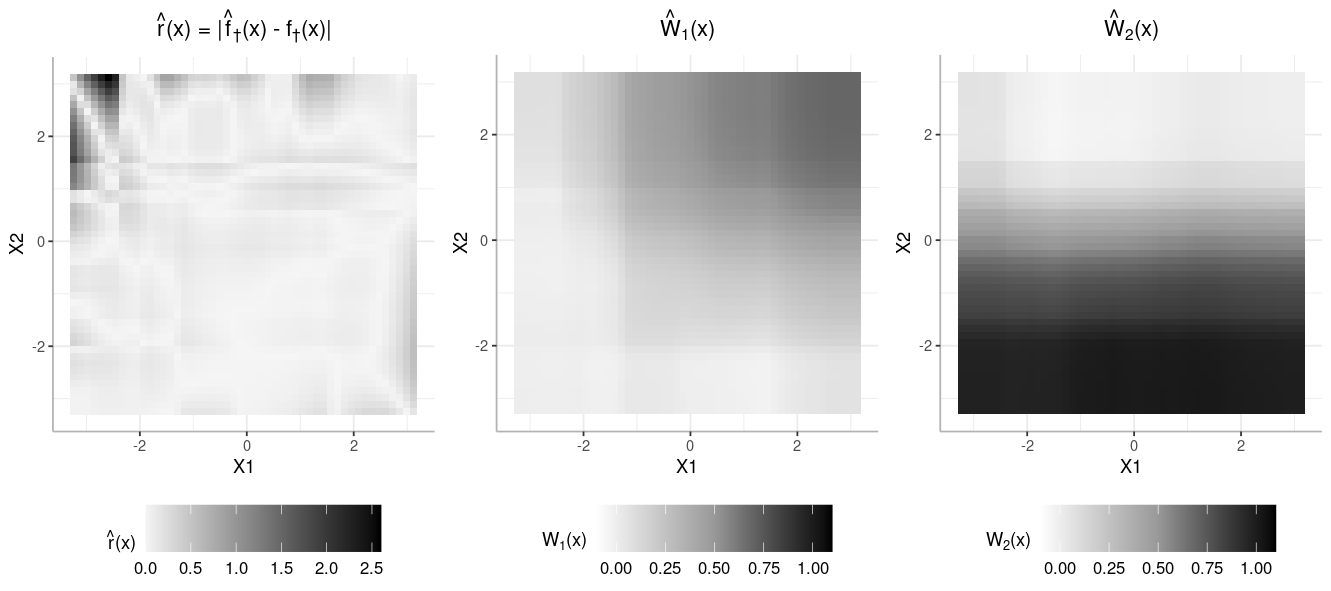}
    \caption{(Left) The mean difference between the predicted system $f_\dagger(\xvec)$, and the true system $f_\dagger(\xvec)$. (Center) The mean weight function for $h_1(\xvec)$. (Right) The mean weight function for $h_2(\xvec)$.}
    \label{fig:ex2_res}
\end{figure}

The results from a 30-tree BART-BMM model are shown in Figure \ref{fig:ex2_res}. The leftmost plot displays the absolute value of the mean residuals, $|\hat{f}_\dagger(\xvec)-f_\dagger(\xvec)|$ where $\hat{f}_\dagger(\xvec)$ denotes the mean prediction from the BMM model. Based on the residual plot, it appears $f_\dagger(\xvec)$ is adequately recovered across the majority of the domain with an RMSE of 0.2575. As expected, the error in the mean prediction noticeably increases in the upper left corner of the domain, where only two training points are included and both simulators are inaccurate.
    
The second and third plots illustrate the posterior mean weight functions for each simulator. Based on the middle plot, the first simulator has increasing utility as $x_1$ and $x_2$ both increase. This is to be expected, as $h_1(\xvec)$ is composed of two expansions centered about $\pi$. Note, the mean value of $w_1(\xvec)$ does not reach 1 in the upper right corner of the domain because the simulator slightly overestimates the peak of $f_\dagger(\xvec)$ in this region. Meanwhile, the posterior mean of $w_2(\xvec)$ indicates $h_2(\xvec)$ has high utility for $\xvec \in [-\pi,\pi]\times[-\pi,0]$, which is to be expected given the nature of the expansions included in this simulator. Moreover, the predictions from $h_2(\xvec)$ appear to align closely with the data, and thus $f_\dagger(\xvec)$, as is  evident by the weights approaching values near 1 in the bottom half of the domain.

\section{Discussion} 
\label{discussion}

A variety of frequentist and Bayesian approaches are available for model averaging and mixing. Each method involves estimating the overall predictive mean or density based on the individual models. The selection between these two objectives should ultimately be guided by the underlying statistical inference one wishes to ascertain. In computer experiments, a primary objective is to recover the underlying system, which is generally expressed as the mean function in an additive model for the observational data. Hence, a mean estimation approach is more desirable when working within this setting compared to a predictive density estimation, which is modeled with the intention of predicting a future observation $\tilde{y}$.  

Example 5.1 compares the proposed mean estimation method versus a density estimation method in Hierarchical Stacking (HS). In HS, the weight functions are learned relative to leave-one-out (LOO) predictive densities under a simplex constraint. These LOO densities incorporate information regarding the mean and variance of each EFT at a given $x$. In portions of the domain where a model may rapidly diverge, the resulting LOO predictive density is shrunk towards 0. In turn, the corresponding weight function will approach 0, however it may struggle to obtain a small enough value to shrink out the effect of the diverging mean. Meanwhile, shrinking the effect of a diverging prediction appears to be easier when mixing the mean predictions from each EFT. 


The primary objective of the weight functions is to re-scale the predictions given by each individual model so that a linear combination of these predictions can adequately recover the true system. Given the prior regularization method applied to the weight functions, exact interpretation of the resulting values can be unclear. 
However, using this regularization perspective, one can conclude that weight functions which fall close to 0 within a particular subregion indicate that the corresponding model is unnecessary for the overall prediction. Meanwhile, a model which is the unique local expert within a particular region should be weighted by values close to 1. Overall, a joint interpretation of the weight functions is appropriate, particularly in regions where the weights concentrate around values away from 0 or 1.  These features are observed across each example. 

\begin{figure}[h]
\includegraphics[width=0.85\textwidth, height = 0.375\textwidth]{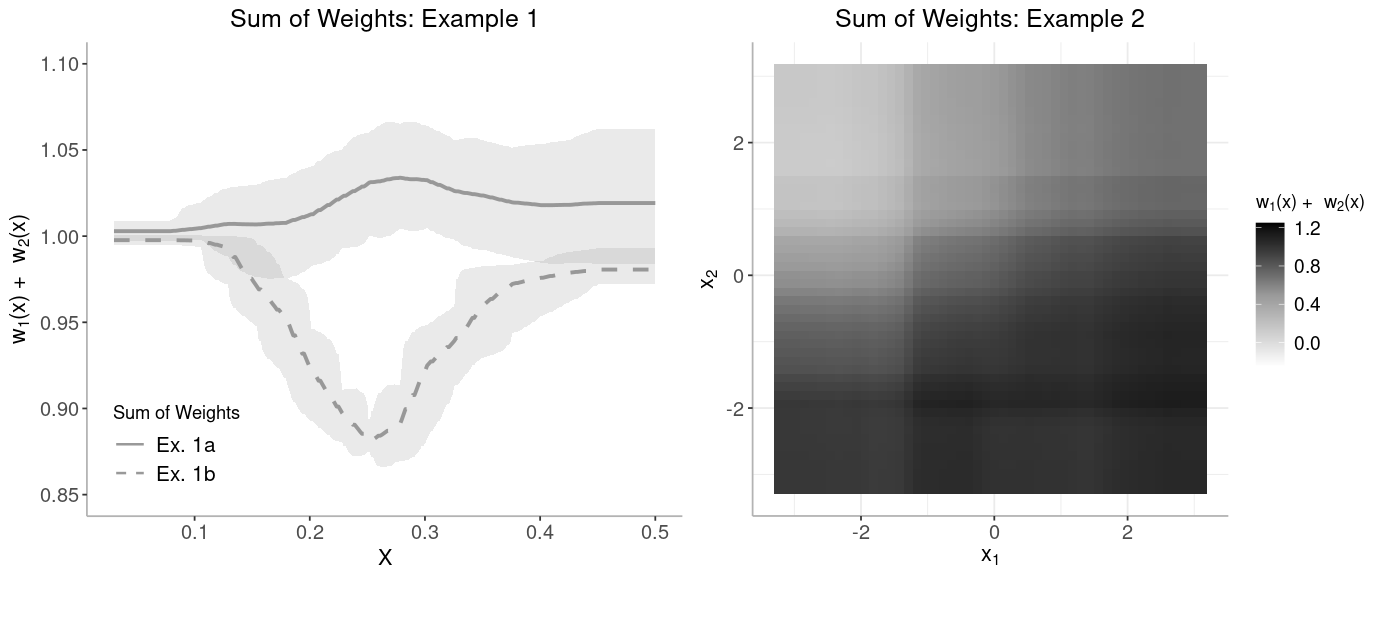}
    \centering
    \caption{(Left) The posterior mean estimates and 95\% credible intervals (shaded) of the sum of weight functions from Examples 1a and 1b (solid and dashed). (Right) The posterior mean estimate of the sum of weight functions in Example 2.}
\label{fig:sum_wts}
\end{figure}

The benefit of the proposed regularization approach can further be understood through the posterior distribution of the sum of the weight functions, $w_\text{sum}(x) = \sum_{l=1}^Kw_l(x)$, as shown in Figure \ref{fig:sum_wts}. The posterior of $w_\text{sum}(x)$ from Example 1a (left panel, solid) is centered very close to 1 with relatively small amounts of uncertainty. This results because: (i) the prior regularization and (ii) $f_\dagger(x)$ lies between the selected EFTs, which indicates a convex combination is appropriate. Even though a sum-to-one property is not strictly imposed, it appears to naturally occur in this situation where an interpolation of the competing models is appropriate. Meanwhile, the posterior of $w_\text{sum}(x)$ from Example 1b (left, dashed) significantly drops below 1 in the intermediate range of the domain because both EFTs overestimate the true system, which renders a convex combination to be inappropriate. Similar features are observed in the 2-dimensional example, as the mean of $w_\text{sum}(x)$ concentrates around 1 in areas where at least one of the simulators aligns well with the true system. Meanwhile, when neither simulator is accurate (i.e. the top left corner) the the mean value of $w_\text{sum}(x)$ is drastically below 1. From these observations, it appears the BART-BMM approach benefits by not imposing strict assumptions, such as a simplex constraint, on the weights.  


Finally, the weight functions can be used to better understand the $\M$-open assumption associated with the model set. An initial confirmation of the $\M$-open setting can be made when the weight functions noticeably change as a function of the inputs. This observation indicates localized performance of each model, hence one can confirm the true system is not contained in the set. If the weight functions are nearly constant, one may also wish to check the posterior of $w_{\text{sum}}(x)$ to see if the sum of the weights is fixated close to 1. Such a case may suggest model averaging with a simplex constraint could also be an appropriate solution. This alone is not enough to confirm or deny the $\M$-open assumption, however it may indicate that the $\M$-complete or $\M$-closed assumptions are possible for the model set. A final case to consider is the situation where a single model receives a weight near 1 while the effects of the competing models are shrunk to 0 across a subregion of the domain. This situation may indicate the model set is $\M$-closed conditional on the subregion of interest despite falling in the $\M$-open case when considering the entire domain.      


In conclusion, this work proposes a Bayesian treed framework to mix predictions from a set of competing models, each of which are intended to explain the physical system across a subregion of the domain. This approach falls within the class of problems referred to as Bayesian model mixing, as input-dependent weights are defined to reflect the localized behavior of each model. The weight functions are modeled using a sum-of-trees and are regularized via a multivariate Gaussian prior. The tree bases coupled with the regularization approach allows for the weights to be learned in a flexible non-parametric manner free of strict constraints. Using the weight functions, predictions from the individual models are mixed via a linear combination. The success of this mixing approach is demonstrated on three examples, each of which considers models with localized predictive performances. Leveraging the localized behavior of the individual models leads to significant improvements in the posterior prediction and uncertainty quantification of $f_\dagger(x)$ and the overall interpretation of the system compared to existing global and local weighting schemes. 

\section*{Acknowledgements}
The work of JCY and RJF work was supported in part by the National Science Foundation under Agreement OAC-2004601.
The work of MTP was supported in part by the National Science Foundation under Agreements DMS-1916231, DMS-1564395, OAC-2004601, and in part by the King Abdullah University of Science and Technology (KAUST) Office of Sponsored Research (OSR) under Award No. OSR-2018-CRG7-3800.3.
The work of TJS was supported in part by the National Science Foundation under Agreement DMS-1564395 (The Ohio State University).

\bibliography{references}{}
\newpage
\section*{Appendix}

Let $\eta_{pj}$ denote the
$\pth$ terminal node in the $\jth$ tree. Without loss of generality, assume $(x_1,y_1),...,(x_{n_p},y_{n_p})$ lie in the hyper-rectangle defined by $\eta_{pj}$. Furthermore, define each residual as 
\[r_i = y_i - \sum_{q\ne j}\fhatvec^\top(\xvec_i)\;\gvec(\xvec_i, T_q, M_q), \ \ i=1,\ldots, n_p
\] These are collected in an $n_p$ dimensional vector $\rvec_{pj} = (r_1,...,r_{n_p})^\top$. Finally, let $\hat{\Fvec}_{pj}$ denote the  $n_p \times K$ matrix whose $\lth$ column is  $(\fvec_l(\xvec_1),...,\fvec_l(\xvec_{n_p}))^\top$. Due to the independence and constant variance assumptions, the model for the vector of residuals along with the associated priors is defined by 
\begin{align*}
    \rvec_{pj} \mid \muvec_{pj}, T_j, \sigma^2 &\sim N_{n_p}\Big(\hat{\Fvec}_{pj}\muvec_{pj}, \sigma^2 I_{n_p} \Big) \\[5 pt]
    \muvec_{pj} \mid T_j &\ind N_K(\betavec_{pj}, \boldsymbol\Sigma) \\[5 pt]
    \sigma^2 &\sim \lambda\nu/\chi^2_\nu
\end{align*}

\noindent where it is assumed $\boldsymbol\Sigma = \tau^2 I_K$. 

\subsection*{The Marginal Likelihood} \label{ml_section}

\noindent The marginal likelihood of the residuals in node $\eta_{pj}$ is defined by 
\begin{align} \label{ml_proof}
L(\rvec_{pj} \mid T_j, \sigma^2) = \int L(\rvec_{pj} \mid T_j,\muvec_{pj} ,\sigma^2)\pi(\muvec_{pj}\mid T_j) \; d\muvec_{pj} 
\end{align} Then, it follows, 
\begin{align*}
    L(\rvec_{pj} \mid T_j, \sigma^2) &= \int (2\pi\sigma^2)^{-n_p/2}\exp\Big(-\frac{1}{2\sigma^2} (\rvec_{pj} - \hat{\Fvec}_{pj}\muvec_{pj})^\top(\rvec_{pj} - \hat{\Fvec}_{pj}\muvec_{pj})\Big) \times \\
    &\quad\quad (2\pi\tau^2)^{-K/2}\exp\Big(-\frac{1}{2\tau^2} (\muvec_{pj} - \betavec_{pj})^\top(\muvec_{pj} - \betavec_{pj})\Big) \;d\muvec_{pj} \\[5 pt]
    &= (2\pi\sigma^2)^{-n_p/2}(2\pi\tau^2)^{-K/2} \times \\
    &\quad\quad\int \Big\{ \exp\Big(-\frac{1}{2\sigma^2} (\rvec^\top_{pj}\rvec_{pj} - 2\muvec_{pj}^\top\hat{\Fvec}^\top_{pj}\rvec_{pj} + \muvec^\top_{pj} \hat{\Fvec}^\top_{pj}\hat{\Fvec}_{pj}\muvec_{pj}\Big) \times \\
    &\qquad \qquad \exp\Big(-\frac{1}{2\tau^2} (\muvec^\top_{pj}\muvec_{pj} - 2\muvec^\top_{pj}\betavec_{pj} + \betavec^\top_{pj}\betavec_{pj})\Big) \;d\muvec_{pj}\Big\} \\[5 pt]
    &= (2\pi\sigma^2)^{-n_p/2}(2\pi\tau^2)^{-K/2} \exp\Big( -\frac{1}{2\sigma^2} \rvec^\top_{pj}\rvec_{pj} - \frac{1}{2\tau^2}\betavec^\top_{pj}\betavec_{pj} \Big) \times \\[3 pt] 
    &\quad \int \exp\bigg(-\frac{1}{2}\muvec^\top_{pj}\Big(\frac{1}{\sigma^2}\hat{\Fvec}^\top_{pj}\hat{\Fvec}_{pj} + \frac{1}{\tau^2}\boldsymbol{I}_K\Big)\muvec_{pj} +\Big(\frac{1}{\tau^2}\betavec_{pj} + \frac{1}{\sigma^2} \hat{\Fvec}_{pj}^\top \rvec_{pj}\Big)^\top \muvec_{pj}\bigg) \; d\muvec_{pj}.
\end{align*}
\noindent Now let $\Amat^{-1} = \frac{1}{\sigma^2}\hat{\Fvec}^\top_{pj}\hat{\Fvec}_{pj} + \frac{1}{\tau^2}I_K$ and $\bvec = \Big(\frac{1}{\tau^2}\betavec_{pj} + \frac{1}{\sigma^2}\hat{\Fvec}^\top_{pj}\rvec_{pj}\Big)$. Substituting these terms into the above expression yields
\begin{align}
    L(\rvec_{pj} \mid T_j, \sigma^2) &= (2\pi\sigma^2)^{-n_p/2}(2\pi\tau^2)^{-K/2} \exp\Big( -\frac{1}{2\sigma^2} \rvec^\top_{pj}\rvec_{pj} - \frac{1}{2\tau^2}\betavec^\top_{pj}\betavec_{pj} \Big) \times \label{ml_integral} \\[3 pt] 
    &\quad \int \exp\bigg(-\frac{1}{2}\muvec^\top_{pj}\Amat^{-1}\muvec_{pj} + \bvec^\top \muvec_{pj} \bigg) \; d\muvec_{pj} \nonumber
\end{align} Using Lemma B.1 from \cite{santner2018design} the integral simplifies as 
\begin{equation}
    \int \exp\bigg(-\frac{1}{2}\muvec^\top_{pj}\Amat^{-1}\muvec_{pj} + \bvec^\top \muvec_{pj} \bigg) \; d\muvec_{pj} = (2\pi)^{K/2} |\Amat|^{1/2}\exp\Big(\frac{1}{2}\bvec^\top \Amat\bvec \Big). \label{ml_lemma}
\end{equation} Then, from (\ref{ml_integral}) and (\ref{ml_lemma}), the marginal likelihood simplifies as
\begin{align*}
    L(\rvec_{pj} \mid T_j, \sigma^2) 
    &= (2\pi\sigma^2)^{-n_p/2}(\tau^2)^{-K/2}|\Amat|^{1/2} \exp\Big( -\frac{1}{2\sigma^2} \rvec^\top_{pj}\rvec_{pj} - \frac{1}{2\tau^2}\betavec^\top_{pj}\betavec_{pj} +\frac{1}{2}\bvec^\top \Amat\bvec \Big). \\ 
    &= (2\pi\sigma^2)^{-n_p/2} (\tau^2)^{-K/2}\bigg|\Big(\frac{1}{\sigma^2}\hat{\Fvec}^\top_{pj}\hat{\Fvec}_{pj} + \frac{1}{\tau^2}\boldsymbol{I}_K\Big)^{-1}\bigg|^{1/2} \\[5 pt]
    &\qquad \times \exp\bigg(-\frac{1}{2}\Big(\frac{1}{\sigma^2}\rvec^\top_{pj}\rvec_{pj} + \frac{1}{\tau^2}\betavec^\top_{pj}\betavec_{pj} - \bvec^\top \Amat \bvec\Big)\bigg) 
\end{align*}

\noindent where $\bvec^\top \Amat \bvec = \Big(\frac{1}{\tau^2}\betavec_{pj} + \frac{1}{\sigma^2}\hat{\Fvec}^\top_{pj}\rvec_{pj}\Big)^\top \Big(\frac{1}{\sigma^2}\hat{\Fvec}^\top_{pj}\hat{\Fvec}_{pj} + \frac{1}{\tau^2}I_K\Big)^{-1}\Big(\frac{1}{\tau^2}\betavec_{pj} + \frac{1}{\sigma^2}\hat{\Fvec}^\top_{pj}\rvec_{pj}\Big).$    

\subsection*{The Full Conditional Distribution of $\muvec_{pj}$}

Now consider the full conditional posterior distribution of the terminal node parameter $\muvec_{pj}$. Using Bayes rule,  
\[
\pi(\muvec_{pj} \mid \rvec_{pj}, T_j, \sigma^2) \propto L(\rvec_{pj} \mid T_j, \muvec_{pj}, \sigma^2)\pi(\muvec_{pj}\mid T_j) 
\]

A conjugate prior is assumed for $\mu_{pj}$, thus the terms in the likelihood and prior can be  rearranged to obtain a Normal kernel for the posterior distribution. This process is summarized below.
\begin{align*}
    \pi(\muvec_{pj} \mid \rvec_{pj}, T_j, \sigma^2) &\propto \exp\Big(-\frac{1}{2\sigma^2} (\rvec_{pj} - \hat{\Fvec}_{pj}\muvec_{pj})^\top(\rvec_{pj} - \hat{\Fvec}_{pj}\muvec_{pj})\Big) \times \\& \qquad \exp\Big(-\frac{1}{2\tau^2} (\muvec_{pj} - \betavec_{pj})^\top(\muvec_{pj} - \betavec_{pj})\Big) \\[5 pt]
    &\propto \exp \Bigg\{-\frac{1}{2} \Bigg(\muvec^\top_{pj}\Big(\frac{1}{\sigma^2}\hat{\Fvec}^\top_{pj}\hat{\Fvec}_{pj} + \frac{1}{\tau^2}I_K \Big)\muvec_{pj}  - 2\muvec_{pj}^\top\Big(\frac{1}{\tau^2}\betavec_{pj} + \frac{1}{\sigma^2}\hat{\Fvec}^\top_{pj}\rvec_{pj}\Big)\Bigg) \Bigg\} \\[5 pt]
    &\propto \exp \Bigg\{-\frac{1}{2} \Bigg( \muvec^\top_{pj}\Amat^{-1}\muvec_{pj} - 2\muvec_{pj}^\top \Amat^{-1}\Amat \bvec \Bigg) \Bigg\}
\end{align*}

\noindent where  $\Amat^{-1} = \frac{1}{\sigma^2}\hat{\Fvec}^\top_{pj}\hat{\Fvec}_{pj} + \frac{1}{\tau^2}I_K$ and $\bvec = \frac{1}{\tau^2}\betavec_{pj} + \frac{1}{\sigma^2}\hat{\Fvec}^\top_{pj}\rvec_{pj}$. The previous expression simplifies as 
\begin{align*}
   \pi(\muvec_{pj} \mid \rvec_{pj}, T_j, \sigma^2) &\propto \exp\Big(-\frac{1}{2}(\muvec_{pj} - A\bvec)^\top \Amat^{-1}(\muvec_{pj} - A\bvec)\Big)
\end{align*}

\noindent This is the kernel of a Multivariate Gaussian distribution with mean $A\bvec$ and covariance matrix $\Amat$. Thus it follows 
\[ \muvec_{pj} \mid \rvec_{pj}, T_j, \sigma^2 \ind N_K\Big(A\bvec, \Amat\Big)\] replacing $\Amat$ and $\bvec$ with their respective definitions implies  
\[ \muvec_{pj} \mid \rvec_{pj}, T_j, \sigma^2 \ind N_K\Bigg(\Big(\frac{1}{\sigma^2}\hat{\Fvec}^\top_{pj}\hat{\Fvec}_{pj} + \frac{1}{\tau^2}I_K \Big)^{-1}\Big(\frac{1}{\tau^2}\betavec_{pj} + \frac{1}{\sigma^2}\hat{\Fvec}^\top_{pj}\rvec_{pj}\Big), \Big(\frac{1}{\sigma^2}\hat{\Fvec}^\top_{pj}\hat{\Fvec}_{pj} + \frac{1}{\tau^2}I_K \Big)^{-1}\Bigg)\]

\subsection*{The Full Conditional Distribution of $\sigma^2$}

\noindent Finally, consider the full conditional posterior for the error variance, which is defined by \[\pi(\sigma^2 \mid \boldsymbol{Y}, T, M) \propto L(\boldsymbol{Y} \mid T, M, \sigma^2)\pi(\sigma^2) \] where $\boldsymbol{Y} = (y_1,...,y_n)^\top$, $T = \{T_1,...,T_m\}$, and $M = \{M_1,...,M_m \}$. \\

\noindent Further, assume a conjugate prior for $\sigma^2$, namely $\sigma^2 \sim \nu\lambda/\chi^2_\nu$ which has a probability density function defined by 
\[ \pi(\sigma^2) = \frac{(\nu/2)^{\nu/2}}{\Gamma(\nu/2)} \lambda^{\nu/2}(\sigma^2)^{-(\nu/2 + 1)}\exp
\Big(-\frac{\nu\lambda}{2\sigma^2} \Big)
\] 

\noindent Due to conjugacy, the full conditional distribution is given by
\begin{align*}
    \pi(\sigma^2 \mid \boldsymbol{Y}, T, M) &\propto (\sigma^2)^{-n/2}\exp\Big\{-\frac{1}{2\sigma^2}\sum_{i = 1}^n \Big(y_i - \fhatvec^\top(\xvec_i)\wvec(\xvec_i)\Big)^2 \Big\}(\sigma^2)^{-(\nu/2+1)}\exp\Big\{-\frac{\nu\lambda}{2\sigma^2} \Big\} \\[5 pt]
    &\propto (\sigma^2)^{-(n/2 + \nu/2 + 1)}\exp\Big\{-\frac{1}{2\sigma^2}\Big(\sum_{i = 1}^n \Big(y_i - \fhatvec^\top(\xvec_i)\wvec(\xvec_i)\Big)^2 + \nu \lambda \Big) \Big\} 
\end{align*}

\noindent This is the kernel of another scaled inverse-$\chi^2$ distribution, namely $\sigma^2 \sim \nu^\prime\lambda^\prime/\chi^2_{\nu^\prime}$ where \[\nu^\prime = n+\nu \quad \text{and} \quad \lambda^\prime = \frac{1}{n+\nu}\Big(\sum_{i = 1}^n \Big(y_i - \fhatvec^\top(\xvec_i)\wvec(\xvec_i)\Big)^2 + \nu \lambda \Big)\] 

\newpage

\section*{Supplementary Material}
\subsection*{An Overview of EFT} \label{eft_overview}

EFTs model physical systems by an infinite expansion of terms  organized in order of decreasing importance according to the power counting principle ~\citep{burgess_2020,doi:10.1142/8619,Georgi:1994qn}. Exact theoretical predictions of the system are obtained by summing over these terms. In practice, only a finite number of lower-order terms are known. Thus, the theoretical prediction can be decomposed using a Taylor-like series which includes the known finite-order expansion along with the induced truncation error. Predictions of experimental quantities can then be represented using an additive model
\begin{align*}
    Y(\xvec) &= f^{(N)}(\xvec) + \epsilon(\xvec) \\
    f^{(N)}(\xvec) &= h^{(N)}(
    \xvec) + \delta^{(N)}(
    \xvec)
\end{align*}
\noindent where $\xvec \in \R^d$ denotes an independent variable associated with the system, $h^{(N)}(\xvec)$ represents the known finite-order expansion of degree $N$, $\delta^{(N)}(\xvec)$ is the associated truncation error, and $\epsilon(\xvec)$ is the random observational error. The accuracy of the finite-order expansion may vary significantly across a subregion of the domain.  For example, a finite-order expansion centered about zero may yield a high-fidelity approximation in the lower regions of the domain. However, the accuracy of the prediction quickly degrades in higher regions of the domain.  

It is further assumed the finite-order expansion can be modeled as a stochastic process. First, the finite-order expansion can factorized as 
\begin{equation} \label{finite_order_exp}
    h^{(N)}(\xvec) = y_{\text{ref}}(\xvec)\sum_{k=0}^Nc_k(\xvec)Q^k(\xvec),
\end{equation}
where $y_{\text{ref}}(\xvec)$ sets the scale of variation,  $c_0(\xvec),...,c_N(\xvec)$ are dimensionless observable coefficients, and $Q(\xvec)$ is a dimensionless expansion parameter. When the scale and expansion parameters are known based on theoretical arguments, the coefficients $c_0(\xvec),...,c_N(\xvec)$ appear to behave as a set of independent and identically distributed curves from a stochastic process  \citep{CTE_EFT}. Thus, a common model for the coefficients is a Gaussian process
\begin{align}
    c_k(\xvec)\mid \boldsymbol{\theta} &\sim GP(\mu, \bar{c}^2r(\xvec,\xvec^\prime;\ell)) \label{coef_model} \\[3 pt] 
    \boldsymbol{\theta} &= (\mu,\bar{c}^2,\ell), \nonumber  
\end{align}

\noindent where $\mu$ denotes a constant mean function and $r(\xvec,\xvec^\prime;\ell)$ represents the covariance function. A common assumption is to set $\mu = 0$, while prior distributions can be assigned to the remaining parameters in the model ~\citep{CTE_EFT}.  

The parameters in (\ref{coef_model}) are learned using a set of evaluations from the kth-order expansion, for $k=0,\ldots,N$, at $n_c$ design inputs $\xvec^{c}_1,\ldots,\xvec^{c}_{n_c}$. Define the set of evaluations of the expansions at the $\ith$ design point by $H(\xvec_i^c) = \{h^{(0)}(\xvec_i^c),\ldots,h^{(N)}(\xvec_i^c) \}$. Given $Q(\xvec)$ and $y_{\text{ref}}(\xvec)$, these evaluations are used to extract the observed finite-order coefficients at each design point $C(\xvec_i^c) = \{c_{0}(\xvec_i^c),\ldots,c_{N}(\xvec_i^c)\}$. A likelihood is formed using $C(\xvec_1^c),\ldots,C(\xvec_{n_c}^c)$ and (\ref{coef_model}). The unknown parameters $\boldsymbol{\theta}$ are estimated through their resulting posterior distributions given these observed coefficients. 


The truncation error accounts for the remaining unknown terms in the series, thus $\delta^{(N)}(\xvec)$ is modeled using a similar factorization 
\begin{equation} \label{discrep_model}
    \delta^{(N)}(\xvec) = y_{\text{ref}}(\xvec)\sum_{k=N+1}^\infty c_k(\xvec)Q^k(\xvec).
\end{equation}
\noindent Using (\ref{coef_model}) and (\ref{discrep_model}) along with properties of the  multivariate normal distributions \citep{ravishanker2021first}, the induced prior on the truncation error term is given by \begin{equation}
    \delta^{(N)}(\xvec) \mid \boldsymbol{\theta}, Q \sim GP\big(m_\delta(\xvec), \bar{c}^2R_\delta(\xvec,\xvec^\prime;\ell)\big), \label{te_prior}
\end{equation}
\noindent with mean and covariance functions
\begin{align}m_\delta(\xvec) &= \mu\;y_{\text{ref}}(\xvec)\;\frac{Q^{N+1}(\xvec)}{1 - Q(\xvec)} \\[3 pt]
R_\delta(\xvec,\xvec^\prime;\ell) &= y_{\text{ref}}(\xvec)y_{\text{ref}}(\xvec^\prime)\;\frac{[Q(\xvec)Q(\xvec^\prime)]^{N+1}}{1 - Q(\xvec)Q(\xvec^\prime)}. \label{te_cov}
\end{align} 
The unknown parameters in (\ref{te_prior}) - (\ref{te_cov}) originate from the coefficient model in (\ref{finite_order_exp}). Thus, the mean and covariance functions which characterize the discrepancy model are also learned using the set of evaluations of the finite-order expansions at the $n_c$ design points. This is a unique property of EFTs, as observational data is not required to learn the model discrepancy.         

When the finite-order expansion is computationally inexpensive to evaluate, the induced prior on the theoretical predictions, $f^{(N)}(\xvec) = h^{(N)}(\xvec) + \delta^{(N)}(\xvec)$ is given by \begin{equation*}
    f^{(N)}(\xvec)\mid \boldsymbol{\theta},Q,\boldsymbol{h}^{(N)} \sim GP\big(m_{\text{th}}(\xvec), \Sigma_{\text{th}}(\xvec,\xvec^\prime) \big),
\end{equation*} 
\noindent where $m_{\text{th}}(\xvec) = h^{(N)}(\xvec) + m_{\delta}(\xvec)$ and $\Sigma_{\text{th}}(\xvec,\xvec^\prime) = \bar{c}^2R_\delta(\xvec,\xvec^\prime;\ell)$. In the expensive case, a GP can be used to emulate the finite-order expansion and is defined by 
\begin{equation*}
    h^{(N)}(\xvec) \mid \boldsymbol{\theta},Q \sim GP\big(m_N(\xvec), \bar{c}^2R_N(\xvec,\xvec^\prime;\ell)\big).
\end{equation*}  The resulting prior on the theoretical prediction is a GP with mean and covariance functions $m_{\text{th}}(\xvec) = m_{N}(\xvec) + m_{\delta}(\xvec)$ and $\Sigma_{\text{th}}(\xvec,\xvec^\prime) = \bar{c}^2R_N(\xvec,\xvec^\prime;\ell) + \bar{c}^2R_\delta(\xvec,\xvec^\prime;\ell)$. In either case, given a set of model runs $H(\xvec_1^c),\ldots,H(\xvec_{n_c}^c)$, one can obtain posterior predictions $\hat{f}^{(N)}(\tilde{\xvec}_1),\ldots,\hat{f}^{(N)}(\tilde{\xvec}_m)$ at new inputs $\tilde{\xvec}_1,\ldots,\tilde{\xvec}_m$.

\end{document}